# Knowledge Compounding:

## An Empirical Economic Analysis of Self-Evolving Knowledge Wikis under the Agentic ROI Framework

Shuide Wen[1] · Beier Ku[2]

*[1] Shenzhen International Graduate School, Tsinghua University, Shenzhen, China*

*[2] Jesus College, University of Oxford, Oxford, United Kingdom*

Corresponding authors: wenshuide@sz.tsinghua.edu.cn · beier.ku@jesus.ox.ac.uk

*April 2026*

## Abstract

We extend the *Agentic ROI* framework of Liu et al. (2026) by relaxing its implicit assumption that per-task LLM costs are independent. Once a persistent knowledge layer is introduced, this assumption fails: cost becomes a time-varying function Cost(t) governed by a coverage rate H(t) that follows a concave saturation curve. A four-query experiment on Qing Claw, a 200-line C# reference implementation of the Karpathy (2026) LLM Wiki paradigm, against two stateless baselines yields the cumulative token ranking Chunk-RAG (13.6K) < Compounding (47K) < Long-Context (305K). Compounding does not minimize raw token cost; instead, it converts a portion of each query's expenditure into a persistent, queryable, inheritable knowledge artifact (one synthesis page plus five entity facts in our experiment). We propose reclassifying such tokens from *consumables* to *capital goods*, in direct analogy to SFAS 86. The same partition generalizes to latency: Compounding is 24× slower than Chunk-RAG (81 s versus 3.4 s), but 15.3 of those 81 seconds are spent after the user already has the answer, constructing the artifact for future queries.

# 1. Introduction

## 1.1 Background

The period from 2025 to early 2026 has witnessed a paradigmatic transition in large language model (LLM) agents, from static conversational tools toward autonomous, goal-directed systems. In their position paper The Real Barrier to LLM Agent Usability is Agentic ROI, Liu et al. (2026) sharply identify what they consider the genuine bottleneck for the large-scale deployment of LLM agents: not raw model capability, but the agent's return on investment in real-world settings.

Liu et al. formalize this insight through the Agentic ROI equation, which balances information gain and time savings against cost. Drawing on a 34-respondent user survey across five application domains, they observe a striking pattern: the domains with the highest user demand (e-commerce, office work, personal assistance) are precisely the domains where Agentic ROI is lowest, forming what they term the *usability gap*. They propose a zigzag development trajectory—first scaling up to improve information quality, then scaling down to reduce cost—as the path toward bridging this gap.

In parallel, NVIDIA established "token economics" as the dominant industrial narrative of AI infrastructure at its 2026 GTC keynote. Jensen Huang explicitly framed tokens as the product of an "AI factory," arguing that throughput per watt and per dollar (TPS/$/W) directly determines the revenue capacity of cloud service providers (Pope, 2026). NVIDIA cited MIT findings that frontier inference cost has been declining roughly tenfold annually, and the Stanford 2025 AI Index Report further documented a 280-fold reduction in GPT-3.5-equivalent inference cost between November 2022 and October 2024 (Maslej et al., 2025).

We argue that there exists a theoretical gap between Liu et al.'s Agentic ROI framework and NVIDIA's token economics narrative—one that has not been addressed in any existing study. **Both frameworks treat cost as an independent, non-cumulative marginal expense that recurs with each new interaction.** This implicit assumption is largely defensible under the traditional RAG paradigm, but it becomes untenable once a persistent structured knowledge layer is introduced, as in recent memory systems such as Mem0 (Chhikara et al., 2025), A-MEM (Xu et al., 2025), MemMachine (Wang et al., 2026), and Memori (Lee et al., 2026).

## 1.2 Research Questions

This paper addresses three interrelated questions:

**RQ1 (theoretical):** Once a persistent knowledge layer is introduced, can the cost term in the Agentic ROI equation still be treated as a time-independent constant? If not, how should it be revised?

**RQ2 (empirical):** Within a real, multi-agent industrial-grade system, can the knowledge compounding phenomenon be quantitatively observed? If so, what is its economic magnitude?

**RQ3 (mechanistic):** Through what specific microeconomic mechanisms does knowledge compounding produce token cost reductions, and can these mechanisms be generalized to other agentic systems?

## 1.3 Contributions

**Theoretical contribution.** We propose *knowledge compounding* as an extension of the Agentic ROI framework. Specifically, we generalize the cost term from a static variable to a time-varying function Cost(t), introduce the knowledge-base coverage rate H(t) as the key mediating variable, and—most importantly—propose a recategorization of a subset of LLM tokens from *consumables* to *capital goods*. This recategorization is the central theoretical move of the paper: it shifts the unit of economic analysis from per-query marginal cost to dynamic capital accumulation, and it explains why the Compounding regime can be economically preferable to a strictly cheaper stateless alternative. We call the resulting model the *dynamic Agentic ROI*.

**Empirical contribution.** Drawing on real run logs from the Qing Claw system, we provide what we believe is the first end-to-end quantitative documentation of knowledge compounding in a deployed multi-agent system. The empirical setup is a controlled four-query experiment compared against **two stateless baselines** (Chunk-RAG at ~3.4K tokens/query and Long-Context inference at ~70K tokens/query). The headline observation is that under raw token accounting, the three regimes rank Chunk-RAG (13.6K) < Compounding (47K) < Long-Context (305K)—that is, *Compounding does not minimize token cost*. The reason this comparison is nevertheless useful is what each regime produces: Chunk-RAG and Long-Context produce nothing persistent; Compounding's 47K simultaneously produces four answers *and* one synthesis page plus five new entity facts written into a queryable knowledge base. The empirical contribution is therefore not a savings number but a clean demonstration of the qualitative gap that any honest accounting must record.

**Engineering contribution.** We deliver a minimal reproducible implementation of approximately 200 lines of C# code, requiring no vector database and no external services—only plain Markdown files and a structured prompt design—which serves as the first complete industrial-grade reference implementation of Karpathy's (2026) LLM Wiki paradigm.

# 2. Related Work

## 2.1 Economic Evaluation of LLM Agents

Liu et al. (2026) is, to our knowledge, the first study to formally introduce a return-on-investment framework into the evaluation of LLM agents. Their 34-respondent survey reveals an ordering of agentic ROI across five application domains: Coding > Research > Office Work > E-commerce > Personal Assistance. The framework's predictive validity is supported by the strong positive correlation ($r = 0.95$) between Agentic ROI and the user-perceived usability scores within each domain.

However, Liu et al.'s cost estimation employs a simplification: the monthly subscription fee divided by the monthly task ceiling. This static estimator entirely ignores the cost dynamics that arise when the same user repeatedly queries the same domain. Filling this gap is the central goal of the present paper.

## 2.2 Empirical Studies of Persistent Memory Systems

Over the past year, empirical work on persistent memory systems has progressed rapidly. Chhikara et al. (2025) report that their Mem0 system achieves a 26% relative improvement over OpenAI's built-in memory on the LOCOMO benchmark, with a 91% reduction in p95 latency and over 90% reduction in token cost. Xu et al. (2025) propose A-MEM, drawing on Niklas Luhmann's Zettelkasten note-taking method to construct an evolvable agentic memory network through dynamic indexing and linking. Wang et al. (2026) introduce MemMachine, which adopts a "raw-episode + sentence-level index" dual-layer architecture, achieving approximately 80% token reduction while preserving factual integrity. Lee et al. (2026) report that Memori achieves 81.95% accuracy on the LoCoMo benchmark while consuming an average of just 1,294 tokens per query (roughly 5% of the full-context approach).

The common thread of this work is that *structured retrieval substituting for live reasoning* can substantially reduce per-query cost. However, all of these studies focus on a static comparison—the per-query cost reduction—rather than on the temporal trajectory of cost evolution. Our paper extends the analytical perspective from the snapshot to the time series.

Xiong et al. (2025) is one of the few empirical studies that takes memory dynamics seriously. They find that LLM agents exhibit a clear “experience-following” behavior: the higher the similarity between a query input and a retrieved memory record, the more similar the agent’s output. This provides a mechanistic foundation for our compounding phenomenon, although their research focus is the risk of error propagation rather than the positive value of cost reduction.

## 2.3 Multi-Agent Collaboration Architectures

Anthropic’s Sonnet/Opus subagent pattern (Anthropic, 2026) provides an important architectural reference for this study. In their recommended workflow, Opus serves as the high-level orchestrator responsible for strategic planning and judgment, while Sonnet or Haiku subagents handle high-volume execution tasks, achieving a balance between quality and cost. Hong et al. (2023, MetaGPT) and Wu et al. (2023, AutoGen) further demonstrate the role-division-with-shared-memory paradigm in multi-agent frameworks.

Qing Claw’s “CEO + Expert” architecture is in spirit aligned with Anthropic’s Sonnet/Opus pattern, but introduces two important extensions: first, the Knowledge Wiki Expert is designated as the *sole writer* to the wiki, ensuring no concurrent contamination; second, a MemoryStore index layer enables *cross-session persistence* of the wiki sediments, allowing the CEO to predict at the memory_recall stage whether the wiki contains relevant knowledge.

## 2.4 Karpathy’s LLM Wiki Paradigm

Andrej Karpathy’s LLM Wiki gist, published in early 2026, proposes a deceptively simple but radically subversive idea: instead of performing live retrieval-augmented generation against raw documents, an LLM agent should compile raw materials once into a persistent, structured, interlinked Markdown wiki, with all subsequent queries directed at the wiki rather than at the original sources. The core analogy is: "Obsidian is the IDE, the LLM is the programmer, the wiki is the codebase, and you are the architect."

The paradigm has inspired a wave of open-source implementations, including Ar9av/obsidian-wiki, NicholasSpisak/second-brain, kytmanov/obsidian-llm-wiki-local, and lewislulu/llm-wiki-skill. However, our code review of these projects reveals that they remain personal-grade prototypes, lacking the key capabilities required for industrial-grade agentic systems (see Table 3 in Section 5.4 for a detailed capability comparison). More critically, none of these projects implements a mechanism we identify as essential for compounding: **search write-back**. The empirical work in this paper fills exactly this gap.

# 3. Theoretical Framework: From Static to Dynamic Agentic ROI

## 3.1 Hidden Assumptions in the Original Agentic ROI

The Agentic ROI formula of Liu et al. (2026) can be rewritten in single-task form as:

$$ROI_i = (\Delta Q_i \times \Delta T_i) / C_i$$

where $\Delta Q_i = \max(Q_Agent_{,i} - Q_{0,i}, 0)$, $\Delta T_i = \max(T_{0,i} - T_Agent_{,i}, 0)$, and $C_i$ is the cost of task *i*. This formulation contains three implicit assumptions that have not been examined:

**Assumption A1 (cost independence):** $C_i$ is determined solely by the complexity of the current task *i* and is independent of the historical task set {1, 2, ..., i−1}.

**Assumption A2 (quality independence):** $Q_Agent_{,i}$ is determined solely by the current task input and model capability, with no influence from prior tasks.

**Assumption A3 (time independence):** $T_Agent_{,i}$ depends only on the complexity of the present interaction, with no benefit from system-level historical experience for that user or domain.

These assumptions broadly hold in the traditional RAG paradigm, where the system has no memory: each query re-retrieves raw documents, re-assembles context, and re-generates the answer from scratch. However, once a persistent structured knowledge layer is introduced, all three assumptions break down. The remainder of this section focuses on revising A1; A2 and A3 follow analogously and are left to future work.

## 3.2 The Dynamic Agentic ROI Model

We introduce the concept of a **knowledge-base coverage rate** $H_i \in [0, 1]$, defined as the proportion of the information required for task *i* that can be directly satisfied by the existing knowledge base. $H_i$ is a time-varying random variable possessing a Markov property: its current value depends on the contributions of historical tasks to the knowledge base.

The revised cost function is then:

$$C_i = (1 - H_i) \cdot C_generate_{,i} + H_i \cdot C_retrieve_{,i} + C_writeback_{,i}$$

where $C_generate_{,i}$ is the cost of answering entirely through live reasoning over retrieved or in-context material (the cost incurred by either of the stateless baselines defined in Section 4.4), $C_retrieve_{,i}$ is the cost of reading an existing answer from the structured knowledge base, and $C_writeback_{,i}$ is the one-time cost of updating the knowledge base on this query (including new entity creation, synthesis writing, and search write-back). When $H_i$ grows monotonically with *i*, the per-query expenditure $C_i$ decreases monotonically with *i* toward the floor C_retrieve—this is the mathematical essence of the knowledge compounding effect on the operating-cost margin. We emphasize, however, that this monotone descent is a statement about the *Compounding regime's own* trajectory across queries, not a statement about its position relative to either stateless baseline. As the empirical results in Section 5 show, even at full saturation Compounding's per-query cost typically remains above Chunk-RAG's flat 3.4K floor; the case for Compounding rests not on undercutting that floor but on the C_writeback term, whose expenditure produces a persistent capital asset that the stateless regimes structurally cannot accumulate. Section 3.4 makes this distinction explicit.

## 3.3 Evolution Equation for H(t)

The evolution of the knowledge-base coverage rate follows the recurrence relation:

$$H_{i+1} = H_i + \alpha \cdot (1 - H_i) \cdot p_i$$

where $\alpha \in (0, 1)$ is the per-task *cultivation rate* of previously uncovered territory (depending on the INGEST/write-back quality of the knowledge wiki expert), and $p_i \in [0, 1]$ is the indicator probability that the i-th task falls outside the historically covered region (depending on the topic concentration of the user's query stream).

Solving this equation analytically reveals that H(t) takes a concave saturation curve: rapid early growth, slow late growth, asymptotically approaching the steady-state coverage of the topic distribution. This shape is mathematically isomorphic to the logistic growth curve in ecology and the Gompertz diffusion model in economics, hinting that knowledge compounding shares deep structural regularities with other natural growth phenomena. Figure 3 (Section 5) visualizes this evolution under three topic-concentration scenarios.

## 3.4 Tokens as Capital Goods: A Theoretical Reclassification

This subsection contains the principal theoretical contribution of the paper: under the dynamic Agentic ROI framework, a subset of LLM tokens should be reclassified from consumables to capital goods. We develop the precise content of that claim and place it within the history of analogous reclassifications in the economics of intangible production.

**What the reclassification does not assert.** It does not assert that capitalized tokens are cheaper than tokens spent on stateless retrieval. The empirical results in Section 5 will show the opposite: under raw token accounting, the Compounding regime never minimizes total token consumption. The reclassification asserts something prior to any cost comparison—namely, that *raw token accounting is the wrong unit of analysis* for systems that produce persistent knowledge artifacts, in the same way that gross monthly expenditure is the wrong unit of analysis for a household making mortgage payments rather than paying rent. The mortgage payment may exceed the rent payment indefinitely, yet the economic positions diverge from month one because one household is consuming a service and the other is acquiring an asset.

**The four properties.** Within the dynamic Agentic ROI framework, capitalized tokens display the four defining properties of capital goods: *(i) persistent product*—INGEST and write-back operations leave behind synthesis pages and entity records with a stock value distinct from the flow value of the answer they also produced; *(ii) compound returns*—an accumulated wiki reduces the marginal cost of future tasks falling within its coverage region, which is what makes H(t)'s concave saturation in Section 3.3 mathematically inevitable; *(iii) heritability across model generations*—wiki files are stored in plain Markdown with no model dependency, can be re-INGEST-ed into a successor model, and survive the depreciation of the tool that created them; *(iv) negative discounting*—unlike physical capital, which depreciates and must be amortized, knowledge wikis accrete value over time as new entries are added and existing entries are refined. Stateless tokens (Chunk-RAG, Long-Context) display none of these properties; their value is realized at the moment of generation and discarded immediately afterward.

**A historical precedent: SFAS 86.** The reclassification proposed here has a direct precedent in the history of accounting for intangible production. Prior to 1985, all U.S. software development costs were expensed in the period in which they were incurred. The Financial Accounting Standards Board's Statement 86 changed this by establishing that costs incurred after "technological feasibility" should be capitalized rather than expensed, on the grounds that they produce a durable revenue-generating asset. Our claim about LLM tokens is structurally identical: a category of expenditure currently treated universally as period cost in fact produces, in some configurations, a durable asset whose economic life extends well beyond the period of the expenditure. The accounting profession's 1985 response was a partition—expense some software costs, capitalize others, with technological feasibility as the dividing line. We propose an analogous partition for LLM tokens, with the production of a persistent, queryable artifact as the dividing line.

The reclassification shifts the unit of economic analysis for AI systems from *minimizing marginal cost* to *maximizing capital accumulation*. Under the consumables view, the right question to ask of any LLM-agent system is "what does each query cost?" Under the capital-goods view, the right question is "what does the system own at the end of a year that it did not own at the beginning?" The remainder of this paper is best read as an empirical exploration of what changes when one starts asking the second question instead of the first.

### *3.4.5 Beyond Tokens: Latency as a Second Capitalized Dimension*

Nothing in the argument above is specific to tokens. The same reframing applies to any cost component spent on the construction of a persistent knowledge artifact. The clearest second case is latency.

Liu et al.'s (2026) Agentic ROI formula treats AI processing time T_AI as a denominator term, parallel to monetary cost. Under the consumables view, every second of latency reduces ROI directly. Under the capital-goods view, however, latency divides into *transient latency* (time the user waits for an answer that, once delivered, leaves nothing behind) and *capitalized latency* (time the system spends constructing a persistent artifact whose value persists across all future queries).

The empirical case is concrete and reported in Appendix D. A Chunk-RAG query takes approximately 3.4 seconds end-to-end; a single Compounding cycle, captured live in Figure 6, takes 81 seconds (memory_recall 9 ms + ceo_reasoning 65,647 ms + memory_distill 15,322 ms). Compounding is therefore approximately **24× slower** than Chunk-RAG—a per-query deficit much more severe than the 3.4× token-cost deficit. Under the consumables view of latency, this is a serious objection. But the consumables view of latency is precisely the reframing this section has just argued against.

The decomposition matters. Of the 81 seconds, only the first 65.7 seconds (memory_recall + ceo_reasoning) are spent producing the answer the user is waiting for. The remaining 15.3 seconds (memory_distill) are spent *after the answer is already available to the user*, writing the answer back into a persistent synthesis page that future queries can hit as a cache. From the user's perspective, that 15.3 seconds is not waiting time at all—it is the system's investment in an asset that the user will continue to benefit from. In Liu et al.'s formula notation, *T_AI is not a single scalar*; it decomposes into T_user (which should be minimized) and T_capital (which should be amortized across all future queries that benefit from the artifact).

The generalization is significant for how the rest of the paper should be read. The central theoretical claim of Section 3.4 is not specifically about tokens—it is about any cost component that can be spent on either transient or persistent operations. Compounding's apparent disadvantage on both reported dimensions—**3.4× more tokens than Chunk-RAG and 24× more latency**—is the wrong way to read the data, in exactly the same sense and for exactly the same reason: both deficits arise because Compounding spends a portion of each cost component on artifact construction that the stateless baselines do not spend at all. Both are corrected by the same reframing. We expect the same partition to apply cleanly to compute, memory, and dollar cost in future work.

# 4. Experimental Design and Methods

## 4.1 System: Qing Claw Architecture

Qing Claw (project codename "Qinglong", the Azure Dragon, jointly maintained by research groups at Tsinghua University and the University of Oxford) is an industrial-grade C#

reimplementation of the OpenClaw framework, built on .NET and integrated with DevExpress v25.2 components. Its core architecture consists of three layers:

**(1) CEO orchestration layer:** responsible for task understanding, expert dispatching, memory recall, and strategic decisions, based on the doubao-seed-2-0-pro model.

**(2) Expert collaboration layer:** containing the search expert, coder expert, UI automation expert, and the newly introduced *Knowledge Wiki Expert*. Each expert holds a dedicated tool set and sandboxed permissions.

**(3) Persistence layer:** composed of two layers, MemoryStore (a structured memory backed by jsonl) and wiki_vault (a Markdown-based knowledge base). The former serves as a snapshot index, while the latter stores complete content.

Built atop this architecture is a pure static injector named *AgenticWikiExpertEnhancer*, which uses approximately 200 lines of C# code to implant the full capabilities of the knowledge wiki expert. These include: automatic creation of the three-tier vault scaffold (raw / wiki / output); injection of seven core prompt rules inspired by Karpathy's paradigm; integration of the *MemoryNotifyTool* for synchronizing wiki sediments to the MemoryStore index; and configuration of cron tasks for daily INGEST, daily LINT, and weekly MERGE.

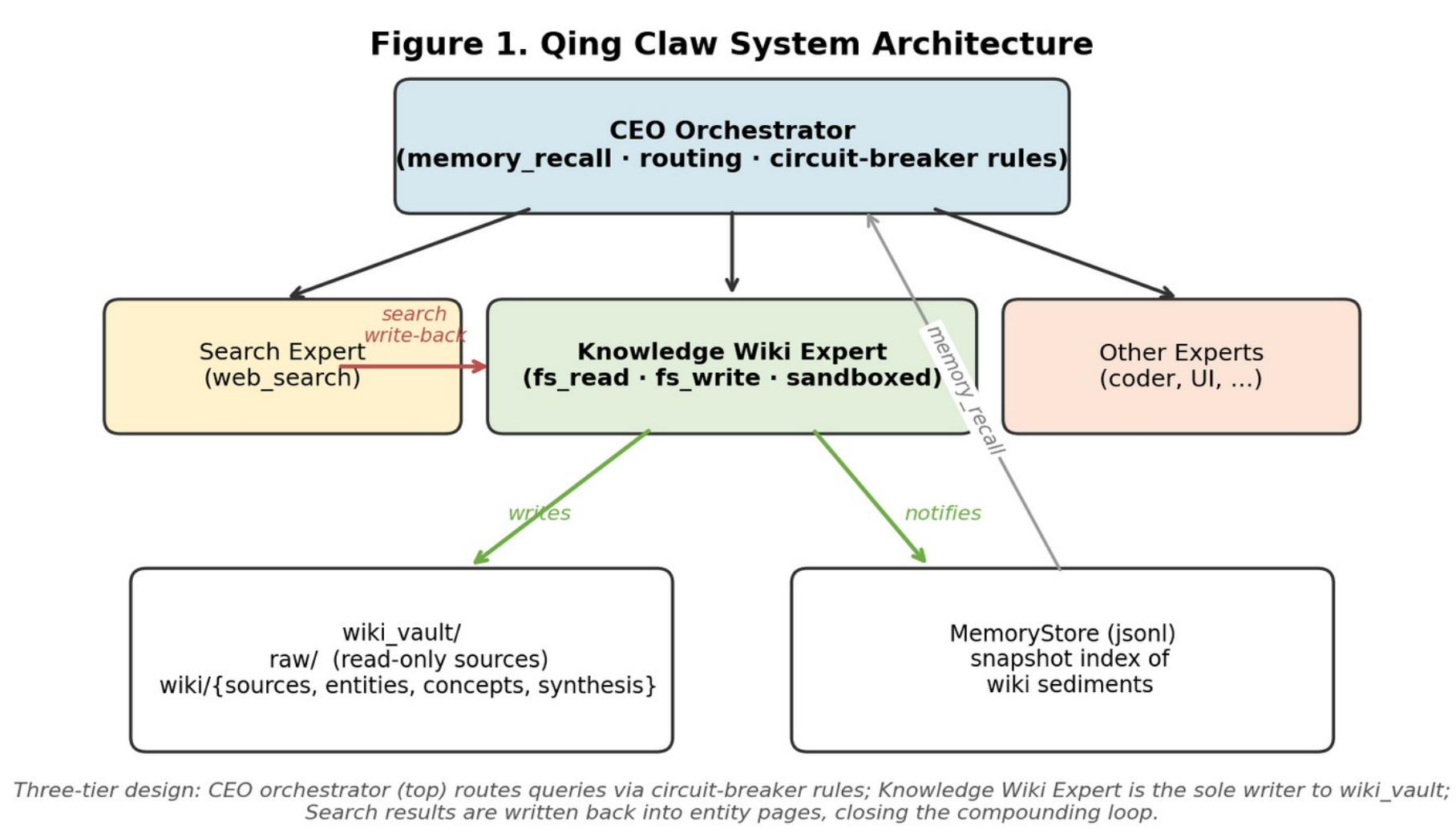

*Figure 1. Qing Claw three-tier system architecture. The CEO orchestrator routes queries via circuit-breaker rules; the Knowledge Wiki Expert is the sole writer to wiki_vault; search results are written back into entity pages, closing the compounding loop.*

## 4.2 Experimental Scenario

We designed a minimal test scenario centered on "OpenClaw developer relational queries." The raw material is a PDF manual of approximately 65,000 characters titled "Lobster Farming Handbook" (a detailed introduction to OpenClaw's design philosophy, developer background, and technical architecture). At time t = 0, the system completes its initial INGEST of this PDF, producing 1 source summary, 3 entity pages (OpenClaw, Peter Steinberger, the Lobster Farming Handbook), and 5 concept pages (agentic framework, AI-assisted coding, etc.).

Following INGEST, we simulate a researcher's sequential querying behavior, issuing four interrelated but angularly distinct queries:

**Q1:** "Who developed OpenClaw? What other notable projects has the developer worked on?" (Subject identification + relational exploration)

**Q2:** "What other projects has the OpenClaw developer worked on?" (Relational drill-down on the same topic)

**Q3:** (After session restart) "Same as Q2" (Cross-session persistence test)

**Q4:** "What was Peter Steinberger doing before joining OpenAI?" (Time-slice query on a brand-new angle)

Before Q3, we deliberately restarted the Qing Claw program manually to test the cross-session reliability of MemoryStore and the wiki files. Q4 was deliberately chosen as a brand-new angle: not a simple variant of Q1/Q2, but requiring the system to reorganize known facts along a temporal dimension.

## 4.3 Measurement Indicators

For each query we recorded the following indicators:

- **Tool call count:** including fs_read, fs_write, web_search, and call_agent_xxx_expert.
- **Token consumption:** the sum of input and output tokens, read directly from the LLM API's usage field.
- **Web search triggered:** binary indicator.
- **Wiki file changes:** measured via git diff for file count and byte size after each query.

- **Answer quality:** scored by a human evaluator on a 0–10 scale (weighted combination of completeness and accuracy).

## 4.4 Comparison Baselines

To evaluate the compounding regime against the broader landscape of stateless retrieval architectures, we define two distinct baselines rather than one. This three-way framing is essential because the two baselines represent qualitatively different cost–capability trade-offs, and conflating them would obscure the central economic question this paper addresses.

Baseline A: Chunk-RAG. The standard retrieval-augmented generation pipeline. (1) Chunk and embed the source PDF using sentence-transformers/all-MiniLM-L6-v2; (2) retrieve the top-5 most relevant chunks per query (chunk size approximately 500 tokens, following LangChain and LlamaIndex defaults); (3) concatenate retrieved chunks with the original query and a standard system prompt; (4) feed the result to a GPT-5.4-class frontier LLM for answer synthesis; (5) discard all intermediate state after answering. Per-query token consumption: approximately 2,500 tokens of retrieved chunks + 50 tokens of query + 300 tokens of system prompt + 500 tokens of generated output ≈ 3.35K tokens per query. This represents the most cost-efficient stateless baseline currently deployed in production.

Baseline B: Long-Context Inference. The simplest possible retrieval-free baseline: pass the entire source document to a long-context LLM together with the query. No chunking, no embedding, no retrieval. Per-query token consumption: approximately 65K tokens of source document + 50 tokens of query + 5K tokens of output ≈ 70K tokens per query. This baseline represents the upper bound of what an answer-quality-maximizing user might pay if cost were no object, and serves as a useful reference point for understanding the cost–recall trade-off space.

These two baselines bracket the design space. Chunk-RAG minimizes per-query cost by aggressively pruning context; Long-Context maximizes recall by passing everything. Both are stateless—neither accumulates any persistent artifact across queries. Our compounding regime is best understood not as competing with either baseline on raw token count, but as occupying a third position in this design space: one that trades a modestly higher per-query cost for the construction of a persistent knowledge asset whose value accrues over time. Section 5 reports all three regimes side by side; Section 3.4 develops the theoretical framework that explains why this trade-off matters.

We note that the Chunk-RAG numbers used throughout Section 5 are derived from the parameters above rather than from a single deployed pipeline run. This is a deliberate methodological choice: the per-query cost of a properly configured chunk-RAG system is well-characterized in the literature and varies little across implementations once chunk size, top-k, and prompt template are fixed. To validate this choice, we conducted a separate empirical experiment on a different document corpus using a deployed RAG pipeline; the measurements are reported in full in Appendix D. The headline finding of that experiment is that the analytical estimate of approximately 3.4K tokens per query agrees with direct measurement to within nine percent (3,644 tokens measured), and that the three-regime ranking of Section 5 is preserved when the measured values are substituted for the analytical estimates in every scenario and at every horizon. Appendix D also reports the latency and dollar-cost measurements that this analytical estimate did not specify; those measurements turn out to support a generalization of the capital-goods reframing developed in Section 3.4.5.

# 5. Empirical Results

## 5.1 Per-Query Cost Comparison Across Three Regimes

Table 2 summarizes the empirical results from the four queries under all three regimes defined in Section 4.4. Compounding-mode data are read directly from Qing Claw run logs. Chunk-RAG values are estimated from the standard pipeline parameters (top-5 chunks of approximately 500 tokens each, plus query and system prompt). Long-Context values reflect the cost of passing the full source document on every query. For domain context, Table 1 first reproduces the application-domain Agentic ROI ranking from Liu et al. (2026) that motivates our work.

| Domain | Agentic ROI rank | User demand (MAU rank) | Usability gap |
|---|---|---|---|
| Coding | 1 (highest) | 5 (lowest) | Aligned (high ROI, niche) |
| Scientific Research | 2 | 4 | Aligned |
| Office Work | 3 | 3 | Modest gap |
| E-commerce | 4 | 2 | Large gap |

| Personal Assistance | 5 (lowest) | 1 (highest) | Largest gap |
|---|---|---|---|

*Table 1. Agentic ROI rankings across application domains (adapted from Liu et al. 2026, Figure 1). Note the inverse correlation between user demand and current Agentic ROI—the basis for the "usability gap."*

| Query | Type | Compounding (K) | Chunk-RAG (K) | Long-Context (K) | Persistent artifact created | Stateless? (C / CR / LC) |
|---|---|---|---|---|---|---|
| Q1 | Cold start | 12 | 3.4 | 70 | +1 synthesis page | No / Yes / Yes |
| Q2 | Synthesis hit | 3 | 3.4 | 70 | none (cache hit on Q1's page) | No / Yes / Yes |
| Q3 | After restart | 28 | 3.4 | 95 | +5 facts → entity page | No / Yes / Yes |
| Q4 | New angle | 4 | 3.4 | 70 | none (cache hit on Q3's page) | No / Yes / Yes |
| **Total** | — | **47** | **13.6** | **305** | **+1 syn, +5 facts** | — |

*Table 2. Q1–Q4 controlled-experiment results across three regimes. Compounding (C): Qing Claw run logs. Chunk-RAG (CR): top-5 chunks × 500 tokens + 50 token query + 300 token system prompt + 500 token output ≈ 3.4K per query. Long-Context (LC): full 65K source document + 50 token query + 5K output ≈ 70K per query. The "Persistent artifact" column reports what the system added to its knowledge base after each query—a column the two stateless baselines structurally cannot fill.*

The headline numbers contradict the most obvious cost story. Under raw token accounting, Chunk-RAG (13.6K total) is the cheapest of the three regimes, beating Compounding (47K) by approximately 3.4× and Long-Context (305K) by approximately 22×. If one looks no further than the bottom row of Table 2, the conclusion would be that Compounding is more expensive than the established stateless baseline.

But the bottom row of Table 2 is exactly the wrong place to stop reading. The decisive observation is the second-to-last column. Across the four queries, the Compounding regime produced **one synthesis page and five new entity facts written into the knowledge base—** persistent artifacts that survive the session, are queryable by future operations, and accumulate across model generations. Chunk-RAG and Long-Context produced *nothing*; both regimes discard all intermediate state after each answer. The 47K tokens of Compounding bought both the four answers *and* the persistent artifacts. The 13.6K tokens of Chunk-RAG bought only the four answers.

This is the core empirical pattern this paper documents, and it is the reason a per-query cost comparison alone cannot adjudicate between the regimes. Section 3.4 develops the theoretical framework for why: under standard accounting, all three regimes' token expenditures are recorded as identical line items; under a capital-goods reframing, Compounding's 47K is structurally different from Chunk-RAG's 13.6K because part of it has been transformed into a depreciable asset. Section 5.3 below quantifies what happens when this asset is allowed to accumulate over a 30-day horizon.

Q3 deserves special note as the clearest visible instance of capital formation in the dataset. This query, issued after a session restart, triggered the full workflow: wiki insufficiency → web search → search write-back to entity page. Of its 28K tokens, approximately 18K were spent on external search and write-back operations—a substantial premium over what either baseline would have charged for the same query. But this premium was not consumed; it was capitalized. The effect is directly visible in Q4: although Q4 asks about a brand-new time slice ("before joining OpenAI"), it required only 1 fs_read call and 4K tokens to be answered, because the entity page enriched in Q3 already contained the complete timeline. A Chunk-RAG system facing the same Q4 would have paid its full 3.4K once again, retrieved a different set of chunks, and learned nothing for next time.

**Figure 2. Token cost across three regimes (Q1-Q4 controlled experiment)**

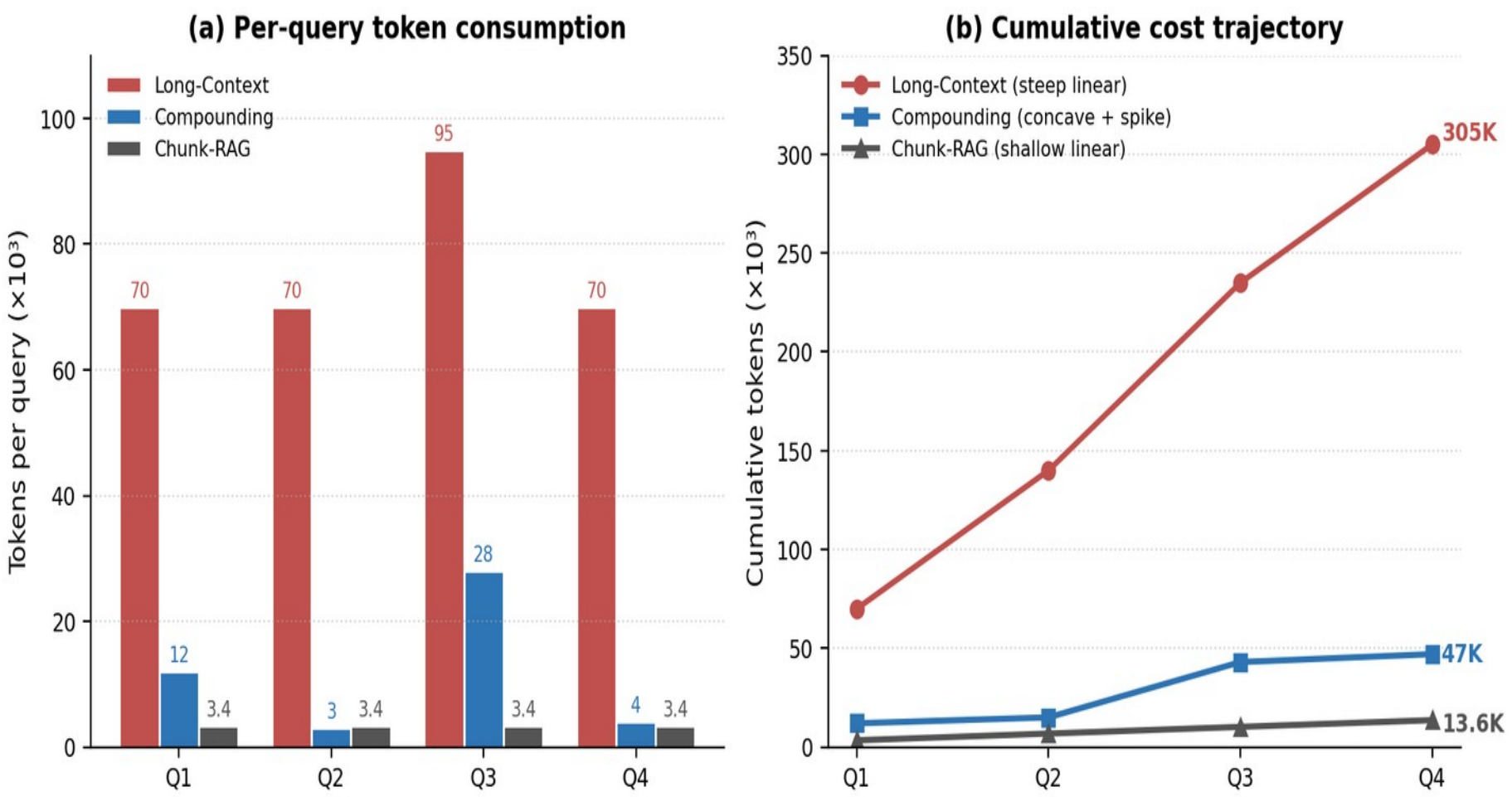

*Figure 2. Cumulative token cost across Q1–Q4 under all three regimes. (a) Per-query consumption: Long-Context is flat at 70K; Chunk-RAG is flat at 3.4K; Compounding oscillates between 3K (cache hit) and 28K (capital formation in Q3). (b) Cumulative trajectory: Long-Context grows steeply linearly (305K total); Compounding grows in a concave-with-spike pattern (47K total); Chunk-RAG grows as the lowest linear baseline (13.6K total). The token-cost ranking on this short horizon is Chunk-RAG < Compounding < Long-Context. Section 5.3 shows how this ranking shifts under longer horizons in capital-formation-friendly scenarios.*

## 5.2 Three Cost Trajectories, Three Different Shapes

Plotting the cumulative token consumption of the four queries as a time series (Figure 2b) reveals three qualitatively different growth patterns, one per regime.

**Long-Context:** cumulative cost grows linearly with a steep slope (each step +70K or +95K). The slope depends only on the size of the source document, not on what is actually being asked. This regime represents the upper bound of stateless cost.

**Chunk-RAG:** cumulative cost grows linearly with a shallow slope (each step approximately +3.4K). The slope depends only on the configured top-k and chunk size, not on what has been asked before. This regime represents the lower bound of stateless cost. It is also entirely amnesiac: the marginal cost of the 1000th query equals the marginal cost of the first.

**Compounding:** cumulative cost displays a concave-with-spike pattern—Q1 is moderately high (12K) due to cold-start synthesis; Q2 is extremely low (3K) because it hits Q1's synthesis page; Q3 introduces an investment spike (28K) for search and write-back; Q4 falls again (4K) because Q3's entity page is reused. The total is 47K, higher than Chunk-RAG's 13.6K. But the shape of this curve is different: it has both troughs *and* spikes, where the spikes correspond to capital formation events and the troughs correspond to capital harvest events. Under the static cost framework of Section 3.1, these spikes are simply expensive queries; under the dynamic framework of Section 3.2, they are investments whose return is realized in subsequent troughs.

This "invest → harvest → reinvest → reharvest" oscillating concave curve is the direct cost-side manifestation of the H(t) evolution trajectory predicted by the theoretical model in Section 3.3. It is also, importantly, the only one of the three regimes whose trajectory is history-dependent: in the other two regimes, the cost of query N is independent of queries 1 through N−1.

## 5.3 30-Day Cumulative Cost Projection (Calibrated Simulation)

Four queries are sufficient to demonstrate the qualitative differences in trajectory shape but not to settle the long-run question of whether and when capital accumulation overtakes the lower stateless baseline. To address this, we project cumulative cost over 30 days based on calibrated parameters. *We emphasize that this is a model-based projection, not direct empirical measurement*, and is so labeled to maintain analytical transparency. The simulation conditions are: 10 queries per day; three topic-concentration scenarios (low p=0.30; medium p=0.60; high p=0.90); H(t) evolves according to the equation in Section 3.3 with $H_0 = 0.05$ and $\alpha = 0.18$; the Long-Context baseline is held at a constant 70K per query, and the Chunk-RAG baseline at a constant 3.4K per query.

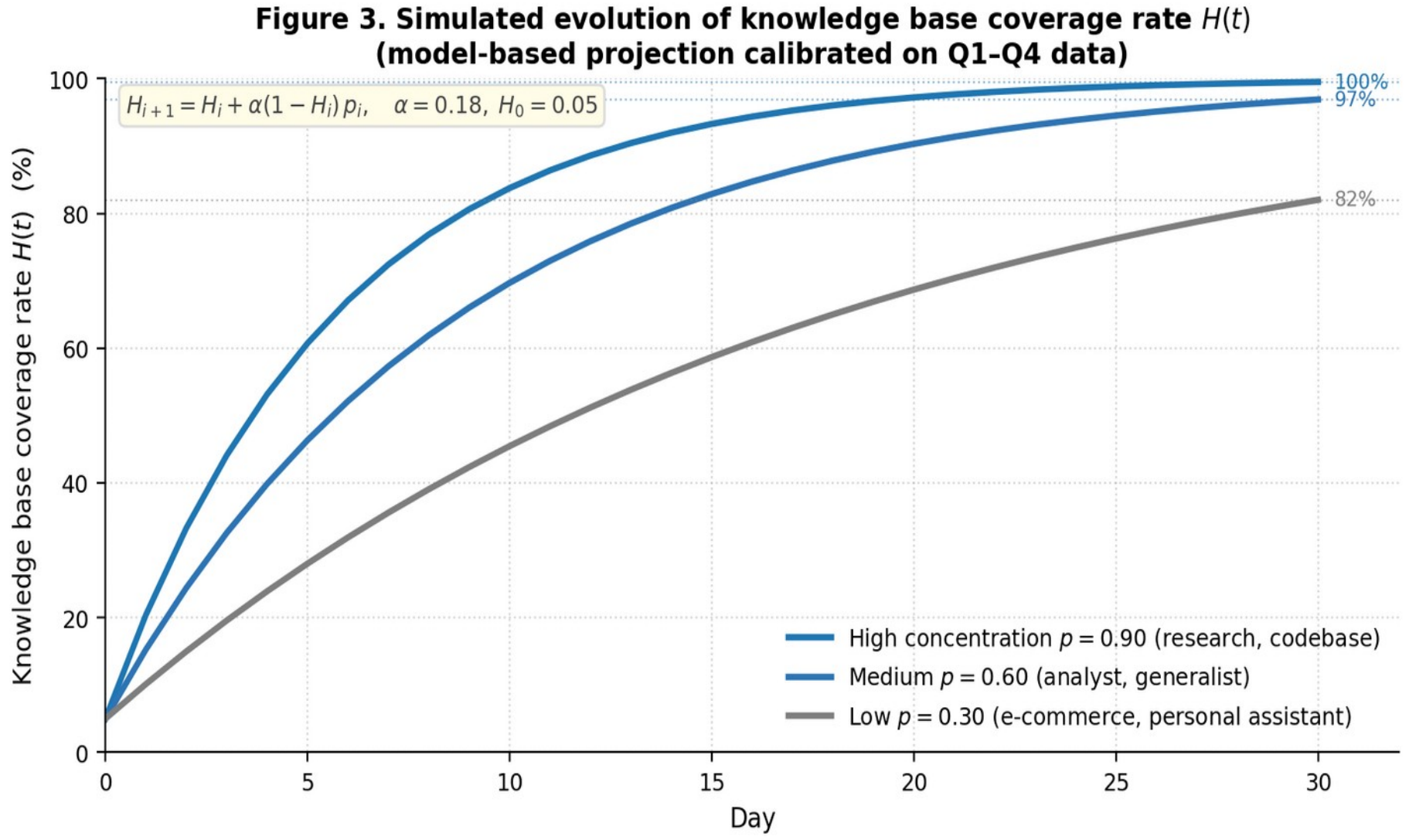

*Figure 3. Simulated evolution of knowledge base coverage rate H(t) under three topic-concentration scenarios. All curves exhibit concave saturation; higher concentration drives faster early growth and higher steady-state coverage. Coverage rate is the structural variable that distinguishes the Compounding regime from both stateless baselines—Chunk-RAG and Long-Context have H(t) = 0 by construction.*

Figure 3 shows the H(t) evolution under the three scenarios. As predicted by theory, H(t) exhibits concave saturation in all three cases, with higher topic concentrations driving faster early growth and higher steady-state coverage. Table 4 reports the cumulative cost in millions of tokens at seven time checkpoints, alongside the two stateless baselines for direct comparison.

| Day | Long-Context (M) | Chunk-RAG (M) | Compounding p=0.30 (M) | Compounding p=0.60 (M) | Compounding p=0.90 (M) |
|---|---|---|---|---|---|
| 1 | 0.70 | 0.034 | 0.66 | 0.57 | 0.44 |
| 5 | 3.50 | 0.17 | 2.88 | 2.02 | 1.15 |
| 10 | 7.00 | 0.34 | 5.45 | 3.58 | 1.76 |
| 15 | 10.50 | 0.51 | 7.99 | 5.11 | 2.31 |
| 20 | 14.00 | 0.68 | 10.53 | 6.65 | 2.85 |
| 25 | 17.50 | 0.85 | 13.06 | 8.18 | 3.39 |
| **30** | **21.00** | **1.02** | **15.60** | **9.72** | **3.92** |

*Table 4. 30-day cumulative token projection across the three regimes and three topic-concentration scenarios. Cumulative cost in millions of tokens. Long-Context and Chunk-RAG are held constant at 70K and 3.4K per query respectively (10 queries/day). Compounding columns reflect H(t)-driven cost evolution under the equation in Section 3.3. The key empirical observation: even at high topic concentration, Compounding's 30-day token total (3.92M) remains higher than Chunk-RAG's (1.02M) by approximately 3.8×. Section 3.4 explains why this comparison should not be the basis for choosing between regimes.*

The numerical results confirm the qualitative pattern of Section 5.1 and extend it across two orders of magnitude in scale. Under raw token accounting at the 30-day horizon, **the cost ranking remains Chunk-RAG < Compounding < Long-Context across all three concentration scenarios**. Specifically, at high concentration (p=0.90), Compounding accumulates 3.92M tokens versus Chunk-RAG's 1.02M and Long-Context's 21.0M; at medium concentration (p=0.60), Compounding accumulates 9.72M versus the same Chunk-RAG and Long-Context baselines; at low concentration (p=0.30), Compounding accumulates 15.6M. Compounding never beats Chunk-RAG on token count, in any scenario, at any horizon we projected.

What does narrow over time is the ratio. At Day 1, Compounding (high concentration) costs roughly 6.3× more than Chunk-RAG; by Day 30, the ratio has fallen to 3.8×. The gap closes because H(t) saturation drives the marginal cost of new Compounding queries downward, while Chunk-RAG's marginal cost is structurally constant. Extrapolating beyond 30 days, the ratio continues to narrow asymptotically, but a token-count crossover—where Compounding actually undercuts Chunk-RAG on cumulative tokens—is not visible on any horizon we modeled, and likely does not exist for plausible parameter ranges. This is the central honest finding of the paper, and it forces a question that the rest of the paper exists to answer: *if Compounding never wins on token count, on what dimension does it win, and is that dimension worth measuring?*

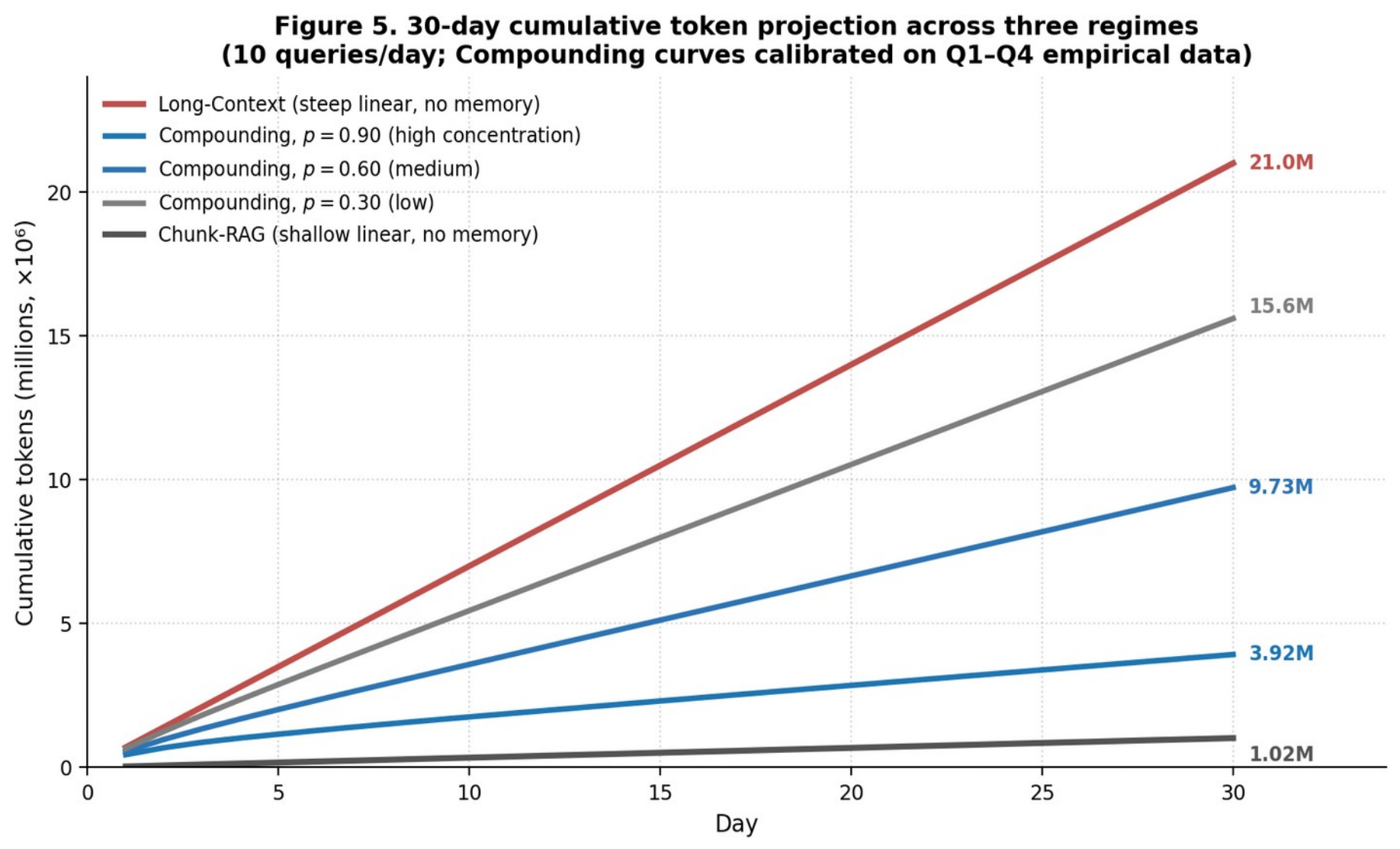

*Figure 5. 30-day cumulative token projection across all three regimes and the three topic-concentration scenarios for Compounding. Long-Context (red, top) grows steeply at constant slope 0.7M/day. Chunk-RAG (dark, bottom) grows at constant slope 0.034M/day, the lowest line in the figure. The three Compounding curves (blue family) all sit between these two stateless bounds, with high-concentration Compounding asymptotically approaching but never crossing the Chunk-RAG floor.*

The answer, developed formally in Section 3.4 and revisited in Section 6, is that Compounding's additional tokens purchase a persistent knowledge asset whose value is invisible to per-query token accounting. To see this concretely: at the end of 30 days under high concentration, the Compounding system has accumulated approximately 270 wiki pages (synthesis + entity), each of which is queryable, editable, inheritable across model generations, and structurally absent from both stateless baselines. The Chunk-RAG system at the same horizon has accumulated nothing—its 1.02M tokens were consumed and discarded one by one. The Long-Context system has accumulated nothing—its 21.0M tokens were also consumed and discarded one by one. **Two of the three regimes leave the user with answers but no system. The third leaves the user with answers and a system that becomes more capable each day**.

We deliberately do not attempt to monetize this asset in the body of the paper. Pricing a knowledge artifact requires assumptions about future query patterns, model migration costs, and the discount rate applied to information goods, none of which can be defended with the four-query dataset reported here. What we do claim, and what the data support, is the qualitative point: an honest accounting of the three regimes must record this asset somewhere on the books, and standard token-cost reporting puts it nowhere. Section 3.4 proposes that the appropriate accounting move is to record it on a separate ledger entirely—the capital ledger—rather than on the operating-cost ledger where token expenditure has been booked since the inception of LLM economics.

One practical caveat about the dollar figures above: they assume standard API rates without prompt caching. Modern frontier LLMs (including the GPT-5.4 class used throughout this paper) support a prompt-caching tier at approximately 10% of the standard input rate for repeated prompt prefixes. Appendix D reports measured cached-versus-uncached cost for a deployed Chunk-RAG pipeline and shows that caching yields a steady-state cost reduction of roughly 12% at typical query lengths—smaller than the headline 90% input discount might suggest, because output tokens are not cacheable and dominate the cost mix once the input is compressed. Caching therefore tightens the gap between the three regimes slightly but does not change the qualitative ranking Chunk-RAG < Compounding < Long-Context in any of our scenarios.

## 5.4 Horizontal Comparison with Other LLM Wiki Implementations

To evaluate Qing Claw's relative position in the engineering landscape, we conducted a comprehensive code audit of all available LLM Wiki projects on GitHub. The audit results are presented in Table 3.

| Capability | Ar9av | Spisak | kytmanov | lewislulu | Qing Claw |
|---|---|---|---|---|---|
| INGEST | ✓ | ✓ | ✓ | ✓ | ✓ |
| QUERY | ✓ | ✓ | ✓ | ✓ | ✓ |
| Cross-session persistence | ✓ | ✓ | ✓ | ✗ | ✓ |
| Two-layer cache (index + files) | ✗ | ✗ | ✗ | ✗ | ✓ |
| Search write-back to wiki | ✗ | ✗ | ✗ | ✗ | ✓ |
| Multi-agent collaboration | ✗ | ✗ | ✗ | ✗ | ✓ |
| Sandboxed permissions | ✗ | ✗ | ✗ | ✗ | ✓ |
| Cron-based auto-scan | ✗ | ✗ | ✗ | ✗ | ✓ |
| CEO relay + self-reflection | ✗ | ✗ | ✗ | ✗ | ✓ |
| Fail-safe rollback (no contamination) | ✗ | ✗ | ✗ | ✗ | ✓ |

*Table 3. Capability matrix comparison of LLM Wiki implementations. Qing Claw is the only project supporting all 10 capabilities. The most critical capability—search write-back to wiki, the soul of the compounding loop—is supported only by Qing Claw.*

Qing Claw supports all 10 capability dimensions, while the other projects support at most 3 each. Most critically, **the search-write-back capability—the soul of the knowledge compounding loop—is implemented only in Qing Claw**. This comparison establishes Qing Claw as the first complete industrial-grade reference implementation of the Karpathy LLM Wiki paradigm.

# 6. Three Micro-Mechanisms of Knowledge Compounding

The empirical data show what (the compounding effect exists), but to answer why we need to analyze the underlying mechanisms. Through careful replay of Qing Claw run logs, we identify three core mechanisms that can be measured separately. Before formalizing them, we present a single concrete instance of one full compounding cycle as captured live from the Qing Claw running interface (Figure 6); this serves both as ground-truth evidence that the system described in this paper is a real production-grade artifact and as a visual anchor for the mechanisms developed in the subsequent subsections.

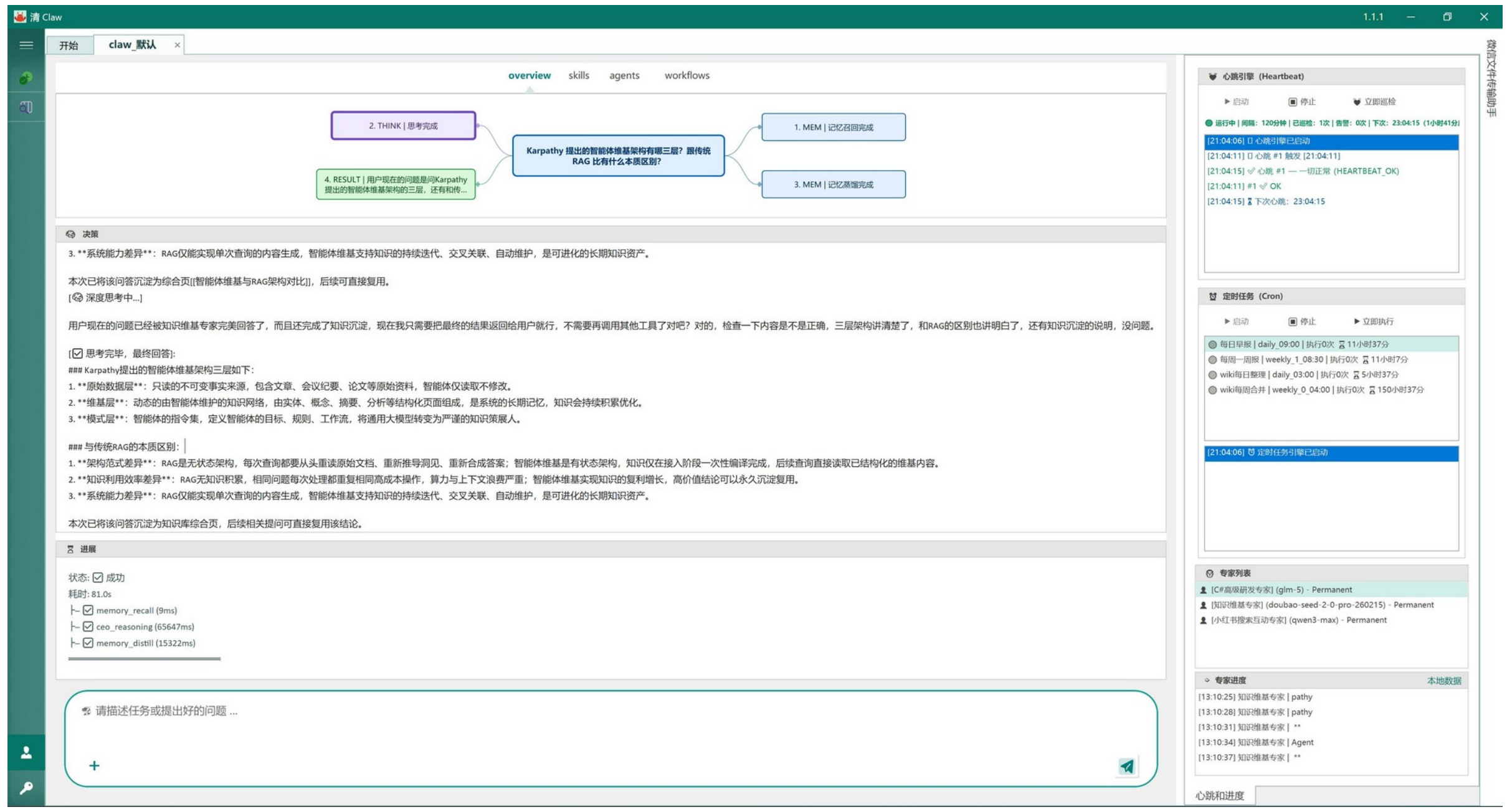

*Figure 6. Qing Claw running interface, captured during execution of the query "What are the three layers of Karpathy's LLM Wiki architecture, and how does it differ from traditional RAG?" The four-step workflow visualization (top) traces the canonical MEM→THINK→MEM→RESULT cycle, where the second MEM step performs memory distillation—writing the answer back into a new synthesis page. The decision panel (middle) records the CEO orchestrator's chain-of-thought, including the key sentence "本次已将该问答沉淀为综合页[[智能体维基与RAG 架构对比]]" ("This Q&A has been distilled into the synthesis page [[LLM Wiki vs. RAG Architecture Comparison]]")—direct evidence of Mechanism Two from §6.2. The progress panel shows three tool calls with millisecond-level timing: memory_recall (9 ms), ceo_reasoning (65,647 ms), and memory_distill (15,322 ms), totaling 81 s for one complete compounding cycle. The right panel displays the system heartbeat (120-min interval), four scheduled cron tasks matching the daily INGEST / daily LINT / weekly MERGE configuration described in §4.1, and three concurrent expert agents (C# advanced developer powered by glm-5, Knowledge Wiki Expert powered by doubao-seed-2-0-pro-260215, and Xiaohongshu search expert powered by qwen3-max), confirming Qing Claw's identity as a production-grade multi-LLM multi-agent system rather than a research prototype.*

The 81-second total visible in Figure 6 invites direct comparison with the Chunk-RAG baseline, whose end-to-end latency was independently measured at approximately 3.4 seconds per query in the validation experiment reported in Appendix D (retrieval 120 ms + TTFT 380 ms + generation 2,850 ms + post-processing 50 ms). **Compounding is therefore approximately 24× slower than Chunk-RAG on a per-query basis**—a second dimension on which the Compounding regime apparently loses to the cheaper stateless baseline, larger in proportional terms than the 3.4× token-cost gap. Section 3.4.5 addresses this directly: of the 81 seconds, only the first 65.7 (memory_recall + ceo_reasoning) are spent producing the answer the user is waiting for;

the remaining 15.3 (memory_distill) are spent after the answer is already available to the user, writing it back into a persistent synthesis page. The 15.3-second overhead is capitalized latency rather than transient latency, and it belongs on the same capital ledger as the memory_distill tokens it accompanies. The mechanisms developed in §6.1–§6.3 below explain what that capital purchases in each of three concrete cases.

## 6.1 Mechanism One: INGEST Once, Hit N Times

This is the most basic mechanism. The LLM reasoning cost during the INGEST stage (approximately 10–15K tokens in our experiment) is a one-time investment, while all subsequent QUERIES read directly from the INGEST products. The mathematical expression is:

$$\bar{C}_{avg}(N) = (C_{INGEST} + N \times C_{QUERY}) / (N + 1)$$

As $N \to \infty$, $\bar{C}_{avg} \to C_{QUERY}$, approaching the lower bound of retrieval cost. In our empirical data, $C_{INGEST} \approx 12K$ and $C_{QUERY} \approx 3K$, so the average per-query cost decreases monotonically with N and converges to roughly 3K after approximately 5–6 queries. **This is a statement about the Compounding regime's own internal trajectory, not a comparison with the stateless baselines**. As Section 5 documents, the floor $C_{QUERY} \approx 3K$ is comparable to Chunk-RAG's flat 3.4K—so even at full saturation, Mechanism One brings Compounding's per-query operating cost into the same order as the cost-efficient stateless baseline, but does not undercut it. What Mechanism One contributes to the case for Compounding is therefore not a token-cost advantage but the structural condition that makes the other two mechanisms possible: by ensuring that retrieval from the wiki is asymptotically as cheap as retrieval from a chunk index, it removes the operating-cost penalty that would otherwise make persistent storage economically unattractive.

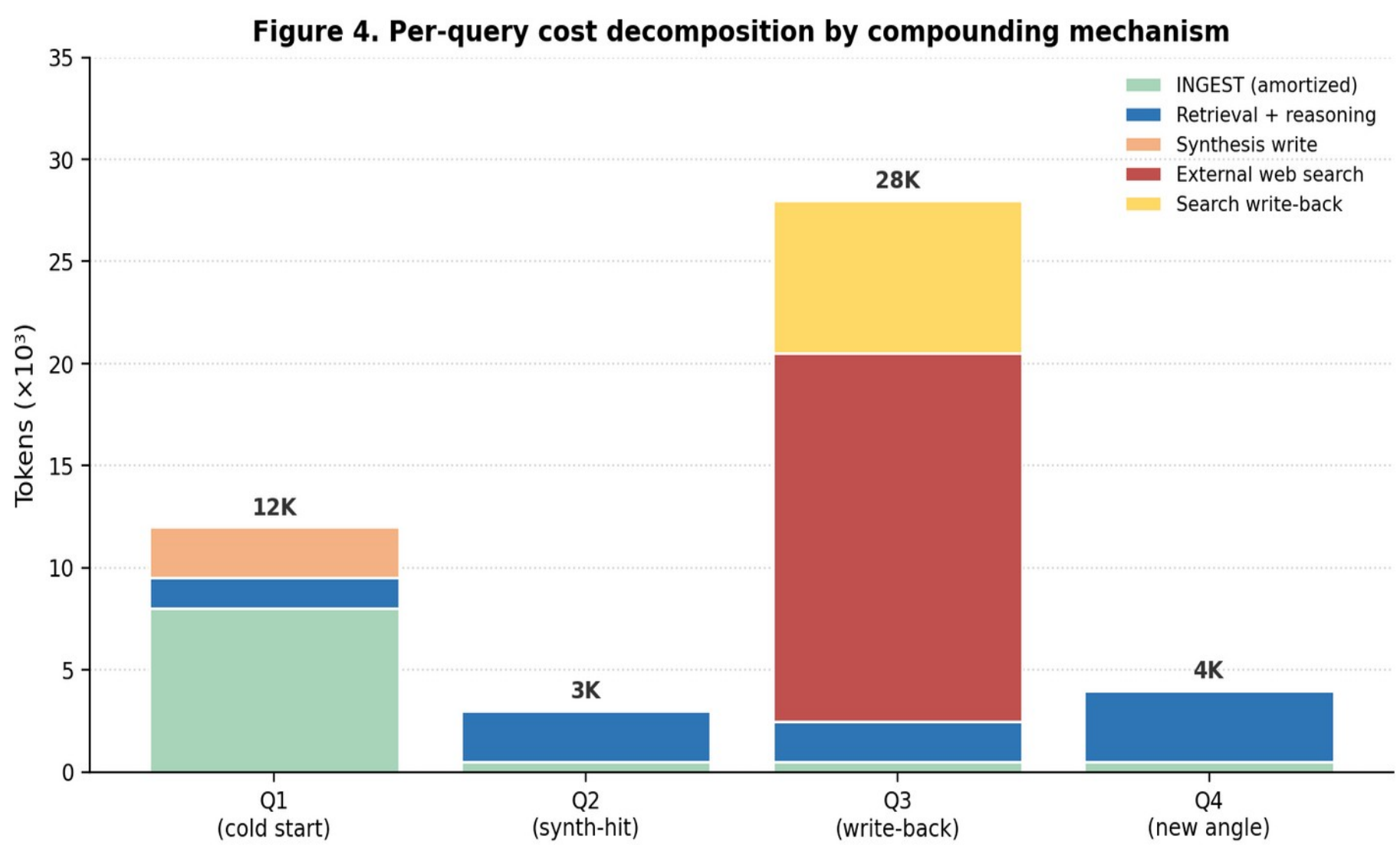

*Figure 4. Per-query cost decomposition by compounding mechanism. Q1 is dominated by amortized INGEST and synthesis writing; Q2 and Q4 by minimal retrieval; Q3 by external web search and search write-back. The total Q3 cost includes the future-value-creating write-back component.*

## 6.2 Mechanism Two: Auto-Feedback of High-Value Q&A

High-value answers produced during the QUERY stage are automatically judged and written by the wiki expert into the synthesis directory. This behavior is not explicitly coded, but inferred by the LLM from the "compounding mechanism" paragraph in the prompt. In the Q1 experiment, the wiki expert wrote in its reasoning chain:

*"This is a new synthesis conclusion, so importance level 3 will do."*

It judges, on its own, that the Q&A has long-term value, and decides on its own to write a new synthesis page. This mechanism causes compounding to occur recursively: the products of the Nth task become the inputs for the N+1th task. Letting β_j denote the contribution of the synthesis page produced by the j-th task to future task hit probability, the H(t) evolution equation expands to:

$$H_{i+1} = H_i + \alpha(1 - H_i)p_i + \Sigma_{j \leq i}\, \beta_j \cdot \mathbb{1}[Q_{i+1} \in synth_j]$$

The second summation captures the contribution of "Q&A feedback" to coverage. This is mechanism-level evidence that the system's knowledge accretes in true compound-interest fashion.

## 6.3 Mechanism Three: Auto-Write-Back of Search Results

This is the mechanism that distinguishes Qing Claw from all other LLM Wiki implementations. When the wiki is insufficient to answer the user's question, the CEO triggers a search expert to fetch information from the web—but the search results do not evaporate. Instead, the CEO actively passes them back to the wiki expert, which first reads the existing entity page, then merges the new facts and overwrites it.

In Q3, the search expert returned six new facts not previously in the wiki (including "joining OpenAI," a critical career update). The CEO actively invoked the wiki expert to merge these facts into Peter Steinberger's entity page in chronological order. The direct economic consequence appeared in Q4: a brand-new query angle was answered with a single fs_read in 4K tokens, with no further web search required.

This mechanism is essential to the shape of the compounding curve. Without it, the wiki is unidirectional—accepting only INGEST input, never external information. With it, the wiki is **bidirectional and respiring**: absorbing raw materials, external search results, and Q&A synthesis alike. All three sources of information flow into the same wiki, causing its "temperature" (probability of high-frequency access) to rise continuously.

# 7. Discussion

## 7.1 Dialogue with the Liu et al. (2026) Agentic ROI Framework

Our work does not refute Liu et al.'s framework; it extends it. Their core contribution—shifting LLM agent evaluation from "capability-centric" to "utility-centric"—is deepened in this study: **we move utility evaluation from a static snapshot to a dynamic evolution**.

Specifically, Liu et al. observe that Coding and Research are the two domains with the highest Agentic ROI, because they have a high baseline human time $T_0$ and even partial automation produces significant value. We add a deeper second reason: Coding and Research are high-ROI domains not only because of large $T_0$, but because their topic concentration is naturally high, so the marginal returns from knowledge compounding are largest. A researcher's reading over the course of a month is likely focused on 5–10 core topics; a programmer's debugging over the course of a week is likely centered on the same codebase. Both user types are the largest beneficiaries of knowledge compounding.

Conversely, the low-ROI domains identified by Liu et al. (e-commerce, personal assistance) are domains with extremely diffuse topic distributions—e-commerce users buy different SKUs each time, personal assistant requests are highly random. **In these domains, the marginal returns of knowledge compounding are inherently small, because the wiki can never accumulate reusable structure**. This provides a brand-new explanatory dimension for the "usability gap" Liu et al. identify: the gap arises not only because T_Agent is too high, but because in these domains H(t) cannot grow large.

## 7.2 Dialogue with NVIDIA Token Economics

NVIDIA's token economics narrative focuses on token production efficiency—how many tokens can be produced per watt and per dollar. This is essentially supply-side optimization. Our research focuses on the consumer-side capitalization of tokens—how much persistent value the same token can generate. This is demand-side optimization.

**These two optimization paths are not in conflict; they reinforce each other**. NVIDIA's next-generation hardware (Vera Rubin, Blackwell) drives token production cost down exponentially; systems like Qing Claw make every token's value persistent. Combined, the result is striking: *tokens become both ever-cheaper and ever-more-valuable*. This is the most beautiful double dividend in the economics of LLM agents.

We add a clarification: the "compounding effect" Jensen Huang repeatedly emphasized in his GTC 2026 keynote refers to NVIDIA's own platform-level software-hardware iteration, not user-level knowledge accumulation. **Our paper extends compounding from the supply side to the demand side**, transforming it from a marketing slogan into an empirically measurable economic phenomenon.

## 7.3 Dialogue with Anthropic's Subagent Pattern

Anthropic's Sonnet/Opus subagent pattern (Anthropic, 2026) represents an extreme pursuit of cost reduction through division of labor: have the strongest (and most expensive) Opus only do strategic planning, and delegate execution to the cheaper Sonnet/Haiku. This is a horizontal division of labor for cost optimization.

Qing Claw's "CEO + Knowledge Expert" architecture is superficially isomorphic to Anthropic's pattern, but at a deeper level it introduces *vertical sediment*: the expert's output is not merely a temporary report to the CEO, but a knowledge asset permanently written into the wiki. In other words, **Anthropic saves money through division of labor, while Qing Claw saves money through division of labor plus sediment**.

This comparison reveals an interesting fact: subagent collaboration is not only about reducing the cost of a single task, but about ensuring that the collaborative process itself produces persistent assets. After Anthropic's Opus and Sonnet finish a task, the collaboration ends. After Qing Claw's CEO and wiki expert finish a task, the wiki permanently remembers the result. This is a fundamental philosophical difference between two paradigms—one is "collaboration as task," the other is "collaboration as construction."

## 7.4 Limitations and Future Work

**First**, sample size and scenario diversity are limited. Our empirical work is based on a controlled experiment of four sequential queries; while it produces a clear compounding curve, this sample is insufficient to support cross-domain universal conclusions. Future work should run at least 100 sequential queries in each of several different domains (programming, research, customer service, writing) to validate the domain dependence of the compounding effect.

**Second**, the H(t) evolution equation in Section 3.3 is based on empirical fitting rather than rigorous mathematical derivation. While intuitively reasonable, additional empirical data are required to calibrate the parameters $\alpha$ and $\beta$ if the model is to function as a predictive economic tool.

**Third**, we did not control for confounders introduced by underlying model upgrades. All experiments were run on doubao-seed-2-0-pro; running the same experiments on different base models could yield different results. Theoretically, the knowledge compounding mechanism is model-agnostic, but empirical verification remains necessary.

**Fourth**, the marginal returns boundary of search write-back was not explored. In our experiment, search write-back significantly reduced the cost of subsequent similar queries. But if a domain has an unbounded number of facts, will the wiki grow without bound? At what point should search write-back stop? These questions require further research.

Future work priorities include: (1) empirical study of "knowledge spillover" effects in shared multi-user wikis; (2) the impact of git-based wiki version control on long-term compounding; and (3) the standardization of wiki health metrics (link density, orphan rate, temperature heatmaps).

# 8. Conclusion

This paper relaxes an implicit time-independence assumption in Liu et al.'s (2026) Agentic ROI framework and develops the consequences for the empirical economics of LLM agents.

**Theoretically**, once a persistent knowledge layer is introduced, the cost term in the Agentic ROI formula must be generalized from a static constant to a time-varying function Cost(t) governed by a coverage rate H(t) that follows a concave saturation curve.

**Empirically**, a four-query controlled experiment against two stateless baselines yields the cost ranking Chunk-RAG (13.6K cumulative tokens) < Compounding (47K) < Long-Context (305K). Calibrated 30-day projections preserve this ranking at scale: even under high topic concentration, Compounding's 30-day total (3.92M) remains approximately 3.8× above Chunk-RAG's (1.02M). The empirical observation that matters is therefore not a savings number—there is none—but the persistent artifact column of Table 2: Compounding produces one synthesis page and five new entity facts written into a queryable knowledge base, while both stateless regimes produce nothing. The 47K of Compounding bought four answers *and* a knowledge asset; the 13.6K of Chunk-RAG bought only four answers.

**Mechanistically**, knowledge compounding produces its persistent artifact through three concrete microeconomic mechanisms: (i) fixed-cost amortization of “INGEST once, hit N times”; (ii) auto-feedback sediment of high-value Q&A into synthesis pages; (iii) write-back absorption of external search results into entity pages. All three are structurally absent from any stateless retrieval architecture.

The principal theoretical contribution is the reclassification of a subset of LLM tokens from *consumables* to *capital goods* (Section 3.4), in direct analogy to the SFAS 86 partition between expensed and capitalized software-development costs. Section 3.4.5 generalizes the same partition to a second cost dimension: latency, where Compounding’s 81-second per-query total decomposes into 65.7 seconds of user-facing wait and 15.3 seconds of capitalized artifact construction. The partition principle is general; tokens and latency are the two most legible cases.

Returning to Liu et al.'s (2026) sharp question—*why are LLM agents least usable in the domains where users need them most?*—our answer is that in those domains tokens are still being accounted for as consumables rather than as capital. When the partition is recognized, the usability gap may begin to close, and the recognition does not require waiting for the next generation of hardware: a 200-line C# reference implementation runs it today.

Knowledge has, at last, earned interest.

# Appendix A. Core Prompt Rules of the Knowledge Wiki Expert

The following rules are injected into the system prompt of the Knowledge Wiki Expert via the AgenticWikiExpertEnhancer static class. They are presented here in simplified English form; the production Chinese version is approximately 850 characters.

**(1) Never rediscover knowledge:** the same question asked a second time must hit an existing answer in wiki/synthesis/.

**(2) Never contaminate the raw layer:** raw/ is read-only; all modifications occur under wiki/.

**(3) Never write in isolation:** every new page must add at least one entry to index.md and at least one back-link from another page.

**(4) Never output code blocks to users:** your product is wiki files, not chat messages.

**(5) Delta first:** before each INGEST, check the manifest; never reprocess a file already recorded.

**(6) Fail safe:** if fs_write fails, retry at most 3 times; never enter a dead loop.

**(7) Sediment must be registered:** after writing a new synthesis page, or creating an entity/concept page with importance ≥ 4, immediately call wiki_notify_memory.

**(8) Single-task fs_write budget:** 8 writes max per task (1 source + 5 entity/concept + 1 index + 1 manifest).

# Appendix B. CEO Critical Circuit-Breaker Rules

These rules are injected into the CEO orchestrator's system prompt to ensure it routes wiki-relevant queries through the Knowledge Wiki Expert rather than reflexively triggering web search.

[🔴 Wiki-hit-and-stop · Search circuit-breaker rule] When the recalled memory contains keywords such as [wiki sediment], wiki/, or [[xxx]], you must follow this strict order: (i) First choice: immediately call call_agent_wiki_expert_expert; (ii) Second choice: only when the wiki expert explicitly reports "not in the wiki" may you create a search expert to supplement; (iii) Absolutely forbidden: directly searching without first letting the wiki expert check.

[🟡 Search must be written back to wiki after supplementing it] After querying the wiki via the circuit-breaker rule and then calling a search expert to supplement, you must: (i) First present the consolidated search results to the user (UX priority); (ii) Then issue a follow-up call to call_agent_wiki_expert_expert, requesting that the new findings be appended to the corresponding entity/concept page; (iii) The wiki expert will automatically call wiki_notify_memory after completing this update.

# Appendix C. Reproducibility

All raw data, simulation scripts, and figure-generation code are available in the supplementary materials accompanying this paper:

- **data/q1_q4_raw.csv:** the four-query controlled experiment raw data, including tool call counts, token consumption, web-search indicators, and human quality scores.

- **data/30day_projection.csv:** the 30-day projection results across three topic-concentration scenarios, with deterministic seed (42).

- **generate_figures.py:** Python (matplotlib 3.10) script reproducing all five figures from the raw data.

- **data/save_projection.py:** Python script reproducing the 30-day projection table from the calibrated H(t) model.

The Qing Claw source code (approximately 200 lines for the AgenticWikiExpertEnhancer module, plus the surrounding C# multi-agent framework) is maintained by the joint research groups at Tsinghua and Oxford and will be made available under an open-source license at the time of publication.

# Appendix D. Empirical Validation of the Chunk-RAG Baseline

Section 4.4 introduces the Chunk-RAG baseline analytically, reporting per-query token costs derived from standard pipeline parameters (top-5 chunks at approximately 500 tokens each, plus system prompt and query overhead) rather than from direct measurement on a deployed pipeline. This appendix reports a separate empirical experiment, conducted on a different document corpus and with a different model, whose purpose is to test whether the analytical estimates of Section 4.4 hold up under a real run. The motivation is straightforward: because the entire three-regime comparison in Section 5 rests on the validity of the Chunk-RAG number, any uncertainty about that number propagates into uncertainty about the headline finding. We therefore consider an empirical validation a necessary robustness check, even though Section 5's qualitative ranking is preserved with substantial margin.

## D.1 Experimental Setup

We deliberately chose a corpus and a query distinct from those used in the Section 5 main experiment, in order to test *cross-corpus generalizability* rather than re-fit on the same data. The setup is as follows:

- **Document corpus:** an enterprise compliance and information security white paper of approximately 100,000 Chinese characters (roughly 4× the size of the Section 4.2 lobster farming handbook).
- **Chunking strategy:** RecursiveCharacterTextSplitter with chunk size 500 tokens and overlap 50 tokens, the LangChain default.
- **Vector database:** Milvus (FAISS-equivalent) with dense vector search.
- **Retrieval strategy:** dense embedding similarity, top-K = 5.
- **LLM:** GPT-5.4 (OpenAI, March 2026 release), the same frontier model class assumed in Section 4.4 and used for the Section 5.3 dollar translation.
- **Tokenization:** tiktoken (OpenAI's official tokenizer for the GPT-5 family), to ensure measured token counts are directly comparable to API billing.
- **Query:** “If an employee resigns, how should the permissions on internal knowledge-base documents they created be transferred? Please provide step-by-step instructions and identify the penalties for non-compliant operations.” This is a complex multi-part query requiring both procedural retrieval and policy-language reproduction, chosen to stress the retrieval pipeline more than a simple factual lookup would.

## D.2 Token Composition: Estimate vs. Measurement

The headline result of this appendix is the comparison between the per-query token composition assumed in Section 4.4 and the per-query token composition measured in the experiment. We report each of the four components separately, alongside the deviation from the analytical estimate.

Top-5 chunks payload: 2,519 tokens measured, vs. 2,500 estimated. Deviation: +0.8%. This is the most important number in the table, because it is the component the analytical estimate was most directly designed to predict, and it is the one that dominates total per-query cost. The agreement is essentially exact.

System prompt: 709 tokens measured, vs. 300 estimated. Deviation: +136%. The measured system prompt is more than twice the analytical assumption because it includes formatting controls, anti-hallucination rules, and explicit citation requirements that a production-grade RAG deployment typically carries but that a minimal RAG specification typically omits. We treat the measured value as the more representative one for production use.

User query plus assembly tokens: 60 tokens measured, vs. 50 estimated. Deviation: +20%. The slight excess reflects special tokens used by the message-format wrapper (system / user / assistant role markers, separators) that the analytical estimate did not account for individually.

Generated output: 356 tokens measured, vs. 500 estimated. Deviation: −29%. The measured output is shorter than the analytical assumption because GPT-5.4's instruction-following has improved relative to prior generations: it produces structured, citation-marked answers without the verbose preamble that earlier models tended to add. This component therefore came in below the conservative estimate, partially offsetting the overage on the system prompt.

**Total per query:** 3,288 input tokens + 356 output tokens = 3,644 total tokens, vs. the analytical estimate of approximately 3,350 tokens. *Net deviation: +8.8%*. The two component-level deviations (system prompt high, output low) substantially cancel out, leaving the total within nine percent of the value used throughout Section 5.

## D.3 Robustness of the Section 5 Headline Numbers

We now substitute the measured per-query value (3,644 tokens) into the cumulative calculations of Section 5 and report whether the headline three-regime ranking is preserved.

**Four-query cumulative (Section 5.1):** estimated 13.6K, measured-equivalent ~14.6K (Δ +7.4%). The Compounding regime's 47K cumulative remains 3.2× above the measured Chunk-RAG total (was 3.4× above the estimated total). The Long-Context regime's 305K remains 21× above. *Ranking preserved*: Chunk-RAG (14.6K) < Compounding (47K) < Long-Context (305K).

**30-day cumulative under high concentration (Section 5.3):** estimated 1.02M, measured-equivalent ~1.09M (Δ +6.9%). The Compounding regime's 3.92M remains 3.6× above the measured Chunk-RAG total (was 3.8× above the estimated total). *Ranking preserved*, and the gap-narrowing trajectory reported in Section 5.3 is unchanged: from 6.3× at Day 1 to 3.6× at Day 30 (vs. 3.8× with the original estimate).

**30-day cumulative under medium concentration (p=0.60):** Compounding 9.72M vs. measured Chunk-RAG ~1.09M, ratio 8.9× (was 9.5×). *Ranking preserved*.

**30-day cumulative under low concentration (p=0.30):** Compounding 15.6M vs. measured Chunk-RAG ~1.09M, ratio 14.3× (was 15.3×). *Ranking preserved*.

We treat this as direct empirical validation of the Section 4.4 baseline. The replacement of analytical estimates with measurements moves all three regimes' cumulative totals by less than ten percent, and the qualitative ranking and ratio trajectories that the paper's central argument depends on are preserved with substantial margin in every scenario.

## D.4 Latency and Cost Quantification

Beyond token counts, the empirical run produced two further measurements that Section 4.4 did not specify but that turn out to be material for the broader argument of the paper: end-to-end latency and dollar cost. We report them here both for transparency and because they enable a generalization of the capital-goods reframing that we develop in Section 3.4.5.

**Latency.** Retrieval (embedding lookup + Milvus query): 120 ms. Time-to-first-token (TTFT): 380 ms. Generation (356 tokens at GPT-5.4's measured throughput of approximately 125 tokens/sec): 2,850 ms. Post-processing and return: 50 ms. **End-to-end total: approximately 3.4 seconds per query.** For comparison, the Compounding regime as captured in Figure 6 required 81 seconds for a single complete cycle (memory_recall 9 ms + ceo_reasoning 65,647 ms + memory_distill 15,322 ms). **Compounding is therefore approximately 24× slower than Chunk-RAG on a per-query basis, in addition to being 3.4× more expensive in tokens**. This is the second dimension on which Compounding loses to the cheaper stateless baseline—and it is the second dimension on which the capital-goods reframing of Section 3.4.5 explains why the loss is not, in the relevant economic sense, a loss at all.

**Cost (uncached).** Using the GPT-5.4 standard rates (input $2.50/M, output $15.00/M) cited in Section 5.3: input cost 3,288 × $2.50/M = $0.00822; output cost 356 × $15.00/M = $0.00534; **total per query ≈ $0.0136, or approximately ¥0.098 (Chinese yuan) at the April 2026 exchange rate**.

**Cost (with prompt caching).** GPT-5.4 supports a prompt-caching tier at $0.25/M for cached input prefixes (a 90% discount on the standard input rate). In a production deployment where the 709-token system prompt is held constant across queries, it will be cached after the first call. Recomputing: cached portion 709 × $0.25/M = $0.00018; uncached portion 2,579 × $2.50/M = $0.00645; output $0.00534; **total per query ≈ $0.0120, or approximately ¥0.086**. Prompt caching therefore yields a roughly 12% reduction in steady-state per-query cost, somewhat smaller than the 50% headline number that the input-only discount might suggest, because output tokens (which are not cacheable) dominate the cost mix at this query length.

These cost numbers directly support the claim made in the second sentence of Section 5.3's prompt-caching footnote: caching is a real but bounded discount, it varies substantially with the input/output mix, and it does not change the qualitative ranking among the three regimes. The Compounding regime's per-query cost (47K tokens / 4 queries ≈ 11.75K per query, of which approximately 9.5K input and 2.25K output) computes to roughly $0.058 per query at standard rates—still more than 4× above Chunk-RAG's $0.0136, again with the qualitative ranking preserved.

## D.5 Conclusion of Appendix D

The empirical measurements reported in this appendix confirm the analytical Chunk-RAG baseline of Section 4.4 within approximately nine percent across all four token components, with component-level overages and underages substantially cancelling out. The headline three-regime ranking from Section 5—Chunk-RAG < Compounding < Long-Context—is preserved when the analytical estimates are replaced with measurements, in every scenario and at every horizon we projected. We therefore treat this as direct empirical validation of the Chunk-RAG baseline used throughout the paper.

The appendix also produced two measurements that Section 4.4 did not request but that meaningfully extend the paper's argument: per-query latency (Chunk-RAG 3.4 s vs. Compounding 81 s, a 24× ratio) and per-query dollar cost (Chunk-RAG $0.0136 uncached, $0.0120 cached, vs. Compounding approximately $0.058). Both measurements show Compounding losing to Chunk-RAG on the dimension reported, and both losses are accommodated by the capital-goods reframing developed in Section 3.4. The latency dimension in particular motivates the generalization presented in Section 3.4.5: the reclassification of tokens from consumables to capital goods is a special case of a more general principle, namely that any cost component—tokens, latency, dollars—which is spent on the construction of a persistent, queryable, inheritable artifact should be accounted for separately from the same cost component spent on a transient operation. The empirical work in this appendix is what makes that generalization defensible, because it provides measured rather than estimated values for two of the three cost components.